\documentclass{iopart}
\usepackage{epsfig}

\begin{document}

\letter%
{Magnetic ordering in Gd$_2$Sn$_2$O$_7$:  the archetypal 
Heisenberg pyrochlore antiferromagnet}

\author{A~S Wills,$^{1,2}$ M E Zhitomirsky,$^3$ B Canals,$^4$ 
J~P~Sanchez,$^3$ P~Bonville,$^5$ P Dalmas de R\'eotier$^3$ 
and A Yaouanc$^3$}

\address{$^1$Department of Chemistry, University College London,
20 Gordon Street, London, WC1H 0AJ, UK}
\address{$^2$Davy-Faraday Research Laboratory, The Royal Institution of 
Great Britain, London W1S 4BS, UK}
\address{$^3$%
Commissariat \`a l'Energie Atomique, DSM/DRFMC/SPSMS, 38054 Grenoble, France}
\address{$^4$%
Laboratoire Louis N\'eel, CNRS, BP-166, 38042 Grenoble,  France}
\address{$^5$%
Commissariat \`a l'Energie Atomique, DSM/SPEC, 
91191 Gif-sur-Yvette, France}

\begin{abstract}
Low-temperature powder neutron diffraction measurements are 
performed in the ordered magnetic state of the
pyrochlore antiferromagnet Gd$_2$Sn$_2$O$_7$.
Symmetry analysis of the diffraction data
indicates that this compound has
the ground state predicted theoretically for a Heisenberg
pyrochlore antiferromagnet with dipolar interactions. 
The difference in magnetic structures of Gd$_2$Sn$_2$O$_7$ 
and of nominally analogous Gd$_2$Ti$_2$O$_7$ is found to be 
determined by a specific type of third-neighbor  
superexchange interaction on the pyrochlore lattice between spins across 
empty hexagons.
\end{abstract}

\begin{figure}[t]
\centerline{\hbox{\epsfig{figure=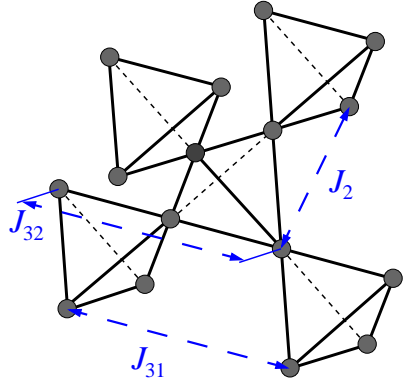,width=0.4\columnwidth}}}
\caption{Pyrochlore lattice of vertex sharing tetrahedra.
Next-neighbor exchanges are shown by long-dashed line.}
\label{pyrolattice}
\end{figure}

Frustration or inability to simultaneously satisfy all 
independent interactions \cite{Anderson} has become an important
theme in condensed matter research, coupling at the fundamental
level a wide range of phenomena, such as high-$T_{\rm c}$
superconductivity, the folding of proteins and neural networks.
Magnetic crystals provide one of the simplest stages
within which to explore the influence of frustration,
particularly when it arises as a consequence of lattice
geometry, rather than due to disorder. For this reason,
geometrically frustrated magnetic materials have been the objects of
intense scrutiny for  over 20
years \cite{Villain}. Particular interest has been focussed on
{\it kagom\'e} and \textit{pyrochlore} 
(see Fig.~\ref{pyrolattice})
geometries of vertex-sharing triangles and tetrahedra
respectively. Model materials with their structures 
display a wide range of exotic low-temperature
physics, such as spin ice \cite{Bramwell_spin_ice}, spin
liquids \cite{YMn2}, topological spin glasses \cite{H3O_Fe}, heavy
fermion \cite{LiV2O4}, and co-operative paramagnetic ground 
states \cite{Tb2Ti2O7}. Research into these
systems was spawned from studies of the archetypal geometrically
frustrated system---the Heisenberg pyrochlore antiferromagnet, which 
in the classical limit was shown theoretically to
possess a disordered ground state. Raju and co-workers found that 
the Heisenberg pyrochlore antiferromagnet with dipolar interactions 
has an infinite number of degenerate spin configurations near 
the mean-field transition temperature, which are described by 
propagation vectors $[hhh]$ \cite{Raju}. Later, Palmer and Chalker 
showed that quartic terms in the free energy lift this degeneracy 
and stabilize a four-sublattice state with the ordering vector 
${\bf k}=(0\,0\,0)$ (the PC state) \cite{Palmer_Chalker}. 

Among various pyrochlore materials 
Gd$_2$Ti$_2$O$_7$ and Gd$_2$Sn$_2$O$_7$ are believed to be 
good realizations of Heisenberg antiferromagnets.
Indeed, the Gd$^{3+}$ ion has a half-filled $4f$-shell with nominally 
no orbital moment.
A strong {\it intrashell} spin-orbit coupling
mixes, however,  $^8S_{7/2}$ and $^6P_{7/2}$ states leading to a 
sizable crystal-field splitting. Recent ESR measurements on dilute systems
gave comparable ratios of the single-ion anisotropy constant $D>0$
to the nearest-neighbor exchange $J$ for the two compounds:
$D/J\sim 0.7$ \cite{Glazkov}. 
This corresponds to a planar anisotropy for the ground state.
Magnetic properties of the stannate and the titanate 
would be expected, therefore, to be very similar.
In this light, the contrasts between the low-temperature behavior of
Gd$_2$Sn$_2$O$_7$ and that of the analogous titanate are 
remarkable. While the titanate displays two magnetic transitions, 
at $\sim 0.7$ and 1~K, to structures with the ordering vector 
${\bf k}=\left(\frac{1}{2}\,\frac{1}{2}\,\frac{1}{2}\right)$  
\cite{Champion,Stewart},
the stannate undergoes a
first-order transition into an ordered state near 
1~K \cite{Bonville}. Furthermore, 
M\"ossbauer measurements indicate
that in the latter the correlated 
Gd$^{3+}$ moments still 
fluctuate as $T\rightarrow 0$~K \cite{Bonville2}.

In this article we demonstrate that Gd$_2$Sn$_2$O$_7$ orders with
the PC ground state, evidencing it as an experimental realization
of a Heisenberg pyrochlore antiferromagnet with dipolar
interactions. We also note that the magnetic structure observed in
Gd$_2$Sn$_2$O$_7$ differs from those observed in the closely
related Gd$_2$Ti$_2$O$_7$, indicating that an extra interaction  is at play in the latter which leads to the observed differences.

\begin{figure}[t]
\centerline{\hbox{\epsfig{figure=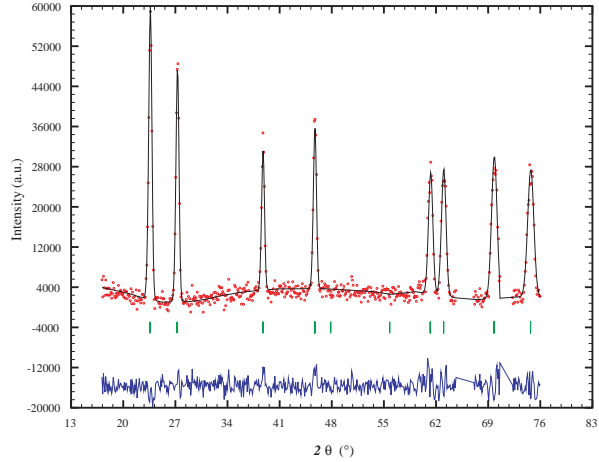,width=0.6\columnwidth}}}
\caption{
Fit to the magnetic diffraction pattern of Gd$_2$Sn$_2$O$_7$ obtained 
from the $\psi_6$ basis state by Rietveld refinement. The dots correspond
to experimental data obtained by subtraction of that measured in
the paramagnetic phase (1.4~K) from that in the magnetically ordered
phase (0.1~K). The solid line corresponds to the theoretical
prediction and the line below to the difference. Positions for the
magnetic reflections are indicated by vertical markers.}
    \label{refinement fig}
\end{figure}

In order to reduce the absorption of neutrons, a 500~mg sample of
Gd$_2$Sn$_2$O$_7$ enriched with $^{160}$Gd was prepared following
conditions given in Ref.~\cite{Bonville}. 
Powder neutron diffraction spectra
were collected with neutrons of wavelength 2.4 \AA\ using the D20
diffractometer of the ILL at two temperatures below
$T_{\rm N}$=1~K (0.1~K and 0.7~K), as well as one temperature in the
paramagnetic phase above $T_{\rm N}$ (1.4~K). Due to the high
residual absorption of the Gd sample, extended counting times of 5 hours per temperature were required. The magnetic contribution to diffraction at 0.1~K
could be well isolated from scattering by the cryostat walls and
dilution insert by subtraction of the spectrum at 1.4~K. Symmetry calculations were made using SARA\textit{h} \cite{Sarah} and Rietveld refinement
of the diffraction data  using
Fullprof \cite{Fullprof}  together with
SARA\textit{h}.

\begin{figure}[t]
\begin{center}
\includegraphics[width=0.8\columnwidth]{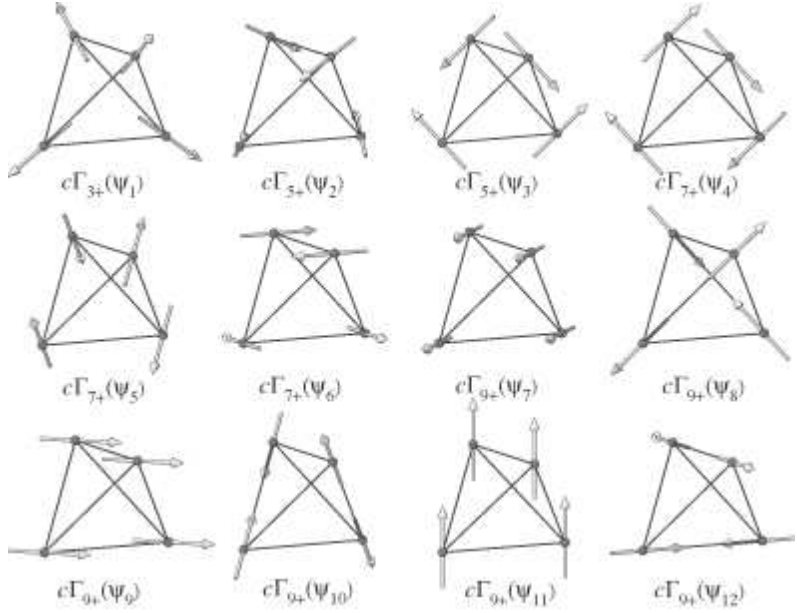}
\end{center}
 \caption{The magnetic structure basis vectors labelled according to the 
different irreducible corepresentations for Gd$_2$Sn$_2$O$_7$.}
    \label{corep fig}
\end{figure}

The magnetic diffraction peaks, Fig.~\ref{refinement fig}, can be 
indexed with the propagation vector ${\bf k}=(0\,0\,0)$.  The 
different types of magnetic structure can be classified in terms of  
the irreducible corepresentations of the reducible magnetic 
corepresentation, $c\Gamma_{mag}$. In the case of Gd$_2$Sn$_2$O$_7$ (space
group $Fd\bar{3}m$ with ${\bf k}=(0\,0\,0)$ and Gd$^{3+}$ ion at the $16d$
crystallographic position) this can be written: 
$c\Gamma_{mag}= 1c\Gamma_{3+} + 1c\Gamma_{5+}
+ 1c\Gamma_{7+} + 2c\Gamma_{9+}$
\cite{Kovalev,corep numbering note}.
Their associated basis vectors are represented in Fig.~\ref{corep fig}. 
Inspection reveals 
that $c\Gamma_{3+}$ 
corresponds to the antiferromagnetic structure observed in
FeF$_3$ \cite{Greedan_FeF3}, 
$c\Gamma_{5+}$ to the linear combination observed in the
model $XY$ pyrochlore antiferromagnet
Er$_2$Ti$_2$O$_7$ \cite{Champion}, $c\Gamma_{7+}$ to the manifold of
states proposed as the ground states for the Heisenberg pyrochlore
antiferromagnet with dipolar terms (the PC ground state), and 
$c\Gamma_{9+}$ to a spin-ice like manifold observed in 
the non-collinear ferromagnetic pyrochlores such as Dy$_2$Ti$_2$O$_7$
\cite{Bramwell_spin_ice}. 
While the phase transition in Gd$_2$Sn$_2$O$_7$ has been shown to be 
first order which allows ordering according to several 
irreducible corepresentations, it is commonly found that 
the terms which drive the transition to being first order are relatively 
weak and cause only minor perturbation to the resultant 
magnetic structure. Following this, we examined whether 
the models detailed above could fit the observed magnetic neutron 
diffraction spectrum. The goodness of fit parameter, $\chi^2$,  for  the fit to models characterised by each irreducible corepresentation are:  $c\Gamma_{3+}$ (69.0), $c\Gamma_{5+}$ (35.6),  $c\Gamma_{7+}$ (5.18),  $c\Gamma_{9+}$ (13.6). We find that the magnetic scattering can only be well 
modeled by $c\Gamma_{7+}$, the PC state in which the  moments of 
a given Gd tetrahedron are parallel to the tetrahedron's edges. 
In this state each moment is fixed to be perpendicular to the 
local 3-fold  axis of each tetrahedron,  consistent with
M\"ossbauer data \cite{Bonville,Bonville2}. 
Powder averaging leads to the structures ascribed to $\psi_4$,
$\psi_5$ and $\psi_6$ being indistinguishable
by neutron diffraction and prevents contributions of the individual basis vectors from
being refined. For this reason only $\psi_6$ was used
in the refinement and the final fit is presented in Figure \ref{refinement fig}. While the  value of the ordered moment,  6(1) $\mu_B$/Gd$^{3+}$, obtained by scaling the
magnetic and nuclear peaks is
imprecise due to the uncertainty over the isotopic
composition of the Gd and the concomitant neutron absorption, it is 
consistent with the free-ion value (7~$\mu_B$) and that measured by M\"ossbauer spectroscopy \cite{Bonville}.

Realization of the PC state in Gd$_2$Sn$_2$O$_7$, 
but not in Gd$_2$Ti$_2$O$_7$, indicates that the magnetic 
Hamiltonian of the titanate contains additional terms.
C\'epas and Shastry \cite{Cepas_Shastry} 
have suggested that next-neighbor exchange 
may stabilize magnetic ordering at
${\bf k}=(\frac{1}{2}\,\frac{1}{2}\,\frac{1}{2})$, though
the corresponding region in the parameter space
was found to be tiny. 
Also, possible exchange paths were not investigated 
in their work  as
both types of third-neighbor exchange (Fig.~1) were
assumed to be equal.

The pyrochlore A$_2$B$_2$O$_7$ structure
has two inequivalent oxygen sites: 
O1 at $(x,\frac{1}{8},\frac{1}{8})$ and 
O2  at $(\frac{3}{8},\frac{3}{8},\frac{3}{8})$.
The oxygen parameter 
is  $x=0.335$ and $0.326$  for
the stannate and the titanate, respectively \cite{structure}.
Using this information we have determined
that the nearest-neighbor gadolinium ions
are connected with short Gd--O1(2)--Gd bonds.
The second-neighbor exchange $J_2$ is produced by two 
distinct Gd--O1--O1--Gd bridges. 
The O1--O1 distance in the first path is $2.63$~\AA\
with two equal bond angles of  $118^\circ$. 
In the second path, $|$O1--O1$|=3.04$~\AA, while the 
angles are $148^\circ$ and $98^\circ$.
(Distances and angles are given 
for Gd$_2$Ti$_2$O$_7$.) 

There are two types of 
third-neighbour pairs in the pyrochlore lattice
indicated in Fig.~1 as $J_{31}$ and $J_{32}$, which  
correspond to three- and two-step Manhattan (city-block) distances, 
respectively. 
The superexchange $J_{31}$ is determined by two equivalent Gd--O1--O1--Gd
paths with $|$O1--O1$|=3.04$~\AA\ and two equal angles  $148^\circ$.
The superexchange $J_{32}$ is produced by
two Gd--O1--O2--Gd bridges with
a significantly larger interoxygen  distance $|$O1--O2$|=3.62$~\AA\ and 
bond angles of $152^\circ$ and $143^\circ$. 
As a result, the two exchange constants for third-neighbor pairs
have to be different with $J_{32}\ll J_{31}$.
The Goodenough-Kanamori-Anderson rules also suggest that 
the second-neighbor constant
$J_2$ must be smaller than $J_{31}$ since 
bond angles in the corresponding superexchange paths 
are significantly closer to $90^\circ$.
Similar relations should hold for next-neighbor
exchange constants in Gd$_2$Sn$_2$O$_7$ with, perhaps,
a somewhat larger ratio $J_2/J_{31}$ due to a larger angle $126^\circ$
in the second-neighbor superexchange path.
The overall effect of further neighbor exchange is, however,
reduced in the stannate because of a larger lattice constant
$a=10.45$~\AA\ compared to $a=10.17$~\AA\ in the titanate \cite{structure}.

\begin{figure}[t]
\begin{center}
\includegraphics[width=0.6\columnwidth]{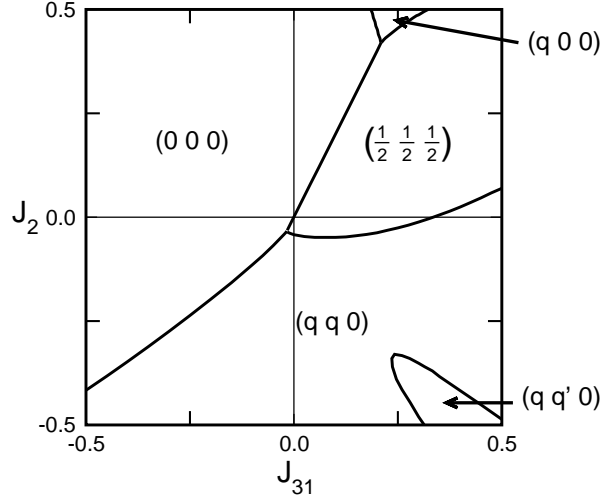}
\end{center}
\caption{Instability wave-vectors for different values of second-
and third-neighbor exchange constants for
a Heisenberg pyrochlore antiferromagnet with dipolar interactions.
Incommensurate states
are indicated by nonzero components of the wave-vectors.
All transition lines are of the first-order.
\label{diagram}}
\end{figure}

Next, we consider the following Hamiltonian

\begin{eqnarray}
\fl \hat{\mathcal H}  =  \sum_{\langle i,j\rangle}
J_{ij}{\bf S}_i\cdot{\bf S}_j + D \sum_i ({\bf n}_i\cdot {\bf S}_i)^2 \label{Hamiltonian} 
+ (g\mu_B)^2 \sum_{\langle i,j\rangle}
\Bigl[ \frac{{\bf S}_i\cdot{\bf S}_j}{r_{ij}^3}
-\frac{3({\bf S}_i\cdot{\bf r}_{ij})({\bf S}_j\cdot{\bf r}_{ij})}
{r_{ij}^5} \Bigr] \ ,
\nonumber
\end{eqnarray}

\noindent
where the superexchange $J_{ij}$ extends up to the third-neighbor pairs of
spins and $D>0$ is a single-ion anisotropy.
The strength of the dipolar interaction between nearest-neighbor
spins  $E_{dd} = (g\mu_B)^2/(a\sqrt{2}/4)^3$ is estimated as
$E_{dd}/J\approx 0.2$ for the titanate \cite{Raju}, where $J$ is
the nearest-neighbor exchange constant (in the following $J\equiv 1$). 
Applying mean-field theory \cite{Raju,Cepas_Shastry} 
and evaluating dipolar sums via the Ewald's summation
we have determined the instability wave-vector for different
values of second- and third-neighbor exchange constants.
Results are essentially independent of the anisotropy constant $D>0$ since the dipolar interaction already selects spins to be orthogonal
to the local trigonal axes ${\bf n}_i$
for the eigenstates with the highest transition temperature.

If only the nearest-neighbor exchange is present, in agreement with 
previous works \cite{Raju,Cepas_Shastry} we find an approximate degeneracy 
of modes along the cube
diagonal with a very shallow minimum $\sim 0.5$\% at
${\bf k}=(\frac{1}{2}\,\frac{1}{2}\,\frac{1}{2})$. 
In such a case, a fluctuation driven
first-order transition is expected to the PC state 
\cite{Palmer_Chalker,first_order}.
The diagram of possible ordering wave-vectors for a restricted
range of $J_{31}$  and $J_2$ are presented in Fig.~4. 
It contains two commensurate states with ${\bf k}=(0\,0\,0)$ and 
${\bf k}=(\frac{1}{2}\,\frac{1}{2}\,\frac{1}{2})$ and three 
incommensurate phases.
Remarkably, already weak {\it antiferromagnetic} $J_{31}$
robustly stabilizes the ${\bf k}=(\frac{1}{2}\,\frac{1}{2}\,\frac{1}{2})$ magnetic
structure, which exists in a wide range
$0<J_{31}<0.335J$. In contrast, a small {\it ferromagnetic}
$J_2$ within a narrow window $-0.04J<J_2<0$is needed to obtain the same ordering without $J_{31}$.
In the whole range of parameters, the eigenstate with
${\bf k}=(\frac{1}{2}\,\frac{1}{2}\,\frac{1}{2})$ 
corresponds to 120$^\circ$ spin structure with $q=0$ in transverse kagom\'e plane 
with zero ordered moment on interstitial sites.
The only remaining degeneracy corresponds to a four-fold orbit of
${\bf k}=(\frac{1}{2}\,\frac{1}{2}\,\frac{1}{2})$ vector.

We have also verified that the second type of third-neighbor exchange
$J_{32}$ does not lead to further stabilization of 
the  $(\frac{1}{2}\,\frac{1}{2}\,\frac{1}{2})$ spin structure. 
Based on these results and the above analysis of the exchange paths
we conclude that it is the third-neighbor exchange across
empty hexagons $J_{31}$ (Fig.~1), which is responsible for
the magnetic structure observed in 
Gd$_2$Ti$_2$O$_7$ \cite{Champion,Stewart}.
In Gd$_2$Sn$_2$O$_7$ further neighbor exchanges play a less
significant role due to a larger lattice constant
and, in addition, the ratio $J_2/J_{31}$ might be enhanced
due to somewhat different bond angles such that
it lies closer to the transition boundary between
${\bf k}=(0\,0\,0)$ and $(\frac{1}{2}\,\frac{1}{2}\,\frac{1}{2})$ states.

In conclusion, the occurrence of the PC state in
Gd$_2$Sn$_2$O$_7$ but not in
Gd$_2$Ti$_2$O$_7$ indicates that the latter possesses additional
contributions, which we identify as a type
of third--neighbor exchange.
Gd$_2$Sn$_2$O$_7$ presents, therefore, the only 
accurate realization of the Heisenberg pyrochlore antiferromagnet
with dipolar interactions.

We are grateful to A. Forget for preparing the $^{160}$Gd enriched sample and to the ILL for provision of neutron time. ASW would like to thank the Royal Society and EPSRC (grant number EP/C534654) for financial support.


\begin{thebibliography}{10}
\bibitem{corep numbering note}
The corepresentations are real and are labelled according to the notation of 
Kovalev \cite{Kovalev} for the parent representation 
and whether the antiunitary 
halfing group was created according to the 
$d(a)=\pm \delta(aa_0^{-1})\beta$, 
where $d(a)$ is the matrix representitive of antiunitary 
symmetry element $a$, 
$\delta(aa_0^{-1})$ is the matrix representitive of unitary symmetry element 
$aa_0^{-1}$, $a_0$ is an antiunitatry generating element and $\beta$ is 
an unitary matrix.
\end{thebibliography}
\end{document}